\documentclass[referee]{aa} 

\usepackage{graphicx}
\begin{document} 
   \title{Influence of the ice structure on the soft UV photochemistry of PAHs embedded in solid water}
   \titlerunning{Influence of ice structure on UV photochemistry of PAH:water ices}

   \author{J.~A. Noble\inst{1}
          \and
          E. Michoulier\inst{2}
          \and
          C. Aupetit\inst{3}
          \and
          J. Mascetti\inst{3}}

   \institute{CNRS, Aix-Marseille Universit\'e, Laboratoire PIIM, Marseille, France.\\
              \email{jennifer.noble@univ-amu.fr}
            \and
             Laboratoire de Chimie et Physique Quantiques LCPQ/IRSAMC, Universit\'e de Toulouse and CNRS, UT3-Paul Sabatier,  118 Route de Narbonne, F-31062 Toulouse, France.
            \and
             Univ. Bordeaux, CNRS, Bordeaux INP, Institut des Sciences Mol\'{e}culaires UMR 5255, F-33405 Talence, France.}

   \date{Received September 15, 1996; accepted March 16, 1997}

 
  \abstract
   {The UV photoreactivity of polycyclic aromatic hydrocarbons (PAHs) in porous amorphous solid water has long been known to form both oxygenated photoproducts and photofragments. }
   {The aim of this study was to examine the influence of the ice structure upon reactivity under soft UV irradiation conditions.}
   {Mixtures of PAHs with amorphous solid water (porous and compact) and crystalline (cubic and hexagonal) ices were prepared in a high vacuum chamber and irradiated using a mercury lamp for up to 2.5 hours.}
   {The results show that the production of oxygenated PAHs is efficient only in amorphous water ice, while fragmentation can occur in both amorphous and crystalline ices. We conclude that the reactivity is driven by PAH-water interactions in favourable geometries, notably where dangling bonds are available at the surface of pores. }
   {These results suggest that the formation of oxygenated PAH molecules is most likely to occur in interstellar environments with porous (or compact) amorphous solid water and that this reactivity could considerably influence the inventory of aromatics in meteorites.}


   \maketitle
%

\section{Introduction}

  In interstellar objects, water ice is typically detected by its broad OH stretching absorption band at 3.1~$\mu$m \citep[e.g.][]{Smith89,Thi06}. In the laboratory, it has been shown that the profile of the 3.1~$\mu$m band can be used to distinguish the four ice structures accessible under the diverse physical conditions of the interstellar medium and planetary systems, namely porous amorphous solid water (pASW), compact amorphous solid water (cASW), cubic crystalline ice (Ic) and hexagonal crystalline ice (Ih), as its profile is sensitive to the ice structure \citep[e.g.][]{Hagen81,Mastrapa09}. 

Although it is generally agreed that it is amorphous upon formation \citep{Hama13}, the precise structure of interstellar ice in different environments remains unclear. Water ice could be expected to exhibit any one of the four structures identified in the lab, or even mixed structural forms, depending on the conditions under which it forms and evolves. If water accretes from the gas phase onto cold (10--20~K) dust grain surfaces in molecular clouds, this will form pASW, but water molecules are initially more likely to form \emph{in-situ} on grain surfaces via radical reactions \citep{Tielens82,Dulieu10}, whose exothermicity may allow restructuring to cASW. Subsequent reactivity on the icy grain surface will also tend to compacify the ice \citep{Accola11}. During the evolution of the icy grain from molecular cloud to core to disk, it will be subject to various types of processing, including heating, which will tend to crystallise the ice \citep{Hagen81}, or particle bombardment \citep[e.g.][]{Baratta91,Dartois15}, which reamorphises. The influence of photon irradiation on ice depends on the initial ice structure, composition and temperature, as well as the photon energy and flux \citep[see e.g.][]{Kouchi90,Oberg16,Tachibana17}. Thus, at later evolutionary stages, the ice structure and ice band profiles depend on its processing history \citep{Boogert15}. In the inner regions around young stellar objects, ice can be crystalline \citep[e.g.][]{Eiroa83,Malfait99}. It is likely that a large fraction of the ice in the outer midplane of protoplanetary disks is amorphous \citep[e.g.][]{Pontoppidan05,Terada07}. This could be prestellar material which remained amorphous due to low temperatures and shielding beyond $\sim$~30~au, or material which has crystallised and reamorphised. The structure of water ice formed by reaccretion following thermal, photo-, or shock-induced desorption will depend on the temperature of the dust grains during accretion. Finally, observations of crystalline ice beyond the snowline and below the photodesorption layer have been attributed to crystallisation induced by grain collisions \citep{McClure15}.
Of course, in none of these environments is interstellar water ice present in its chemically pure form, since other small molecules are formed concurrently or may accrete onto dust grains, being incorporated into their icy mantles. Both the composition and structure is impacted by the inclusion of other molecular species.

Polycyclic aromatic hydrocarbons (PAHs) are believed to represent a significant reservoir of carbon \citep{Tielens08} in the interstellar medium (ISM) in both the gas and solid phases \citep{Salama08}.  Analysis of the aromatic interstellar band (AIB) emission features in the mid infrared (MIR) supports the hypothesis that the interstellar PAH population is centered around molecules of the size C$_{50}$ -- C$_{100}$ \citep{Allamandola89}. PAHs smaller than this are destroyed by UV irradiation \citep{Gordon08} in regions of high photon flux. However, in molecular clouds, such molecules may survive in icy grain mantles where they are somewhat shielded from radiation. It is postulated that, in the spectra of YSOs, up to 9~\% of the unidentified absorption in the 5 -- 8~$\mu$m range can be attributed to neutral PAHs frozen out in dust grain mantles \citep{Hardegree14}.
Much experimental investigation has been dedicated to the photo-processing of PAHs in water ices upon UV irradiation under different conditions. Under high energy VUV irradiation (centered at 121.6~nm and 160~nm), the formation of PAH cations is observed \citep{Bernstein07,Bouwman09,Bouwman10}, as well as a rich photochemistry leading to the formation of alcohols, quinones and ethers \citep{Bernstein99,Bernstein01,Bernstein07,Bouwman11,Cook15,DeBarros17}.
At lower energy UV ( > 235~nm), the photo-processing of pyrene and coronene adsorbed on and embedded in amorphous solid water (ASW) at 10~K also revealed the production of oxygen-containing PAH products \citep{Guennoun11a,Guennoun11b}. This reactivity occurs at energies below the ionisation energy of isolated PAHs (7.4 and 7.2~eV for pyrene and coronene, respectively) and a current key issue is to understand the mechanism of this solid state photochemistry.

One hypothesis is that photoreactivity, even at lower energy, could be ion-mediated \citep{Bouwman09}. Experimental investigations combined with computational studies have shown that the ionisation of PAHs adsorbed on water ice requires about 1.5 to 2.0~eV less energy than in the case of isolated PAHs \citep{Gudipati04,Woon04}, and that the resulting PAH radical cations would be particularly stable over time \citep{Gudipati06}. That could be accounted for by the fate of the electrons released by the ionisation, that would not be free electrons but would attach to H$_2$O molecules or free radicals such as HO$^{\circ}$ formed during ice photolysis \citep{Gudipati06}. 
More recent studies have demonstrated the role of concentration in the reactivity of small PAHs \citep{Cuyulle14,Cook15,DeBarros17}. Low concentrations of PAHs in water ice favour the formation of oxygenated PAHs, rather than erosion processes, that form small molecules such as CO, CO$_2$, and H$_2$CO \citep{Cook15}.
However, we have observed similar photoreactions (\emph{i.e.}, leading to the formation of hydroxy-PAHs) at low energy and in the absence of ice, \emph{i.e.} via the irradiation at > 235~nm of pyrene and coronene isolated in argon matrices in the presence of well-characterised water clusters containing no more than four water molecules \citep{Simon17,Noble17}. It thus appears that a water ice structure is not necessary to obtain oxygen-containing PAHs, and we postulated the presence of a charge-transfer state to explain the formation of oxygen-bearing photoproducts at low UV energy and without ice \citep{Noble17}.
These unexpected results motivated us to run computational studies on the influence of ice and ice structure on the ionisation energy of PAHs \citep{MichoulierAdsorption,MichoulierIP}. We showed that the adsorption of PAHs onto low density amorphous ice (LDA) clearly leads to a decrease of their vertical ionisation potentials (VIPs), but not by more than -0.8~eV, calling into question the ion-mediated route for low energy irradiation. Another important result is that the VIP variation is very dependent on the number of dangling H bonds interacting with the PAH. Their presence favours an increase of the VIP. In contrast, the interaction of an O at the ice surface with the PAH (and in particular with the H of the PAH, present when the PAH is adsorbed at an angle to the surface), leads to a lowering of the VIP \citep{MichoulierIP}. The differences between PAH interaction with LDA and with Ih can be clearly traced to PAH-surface bonding, depending on the orientation of water molecules at the ice surface.
It has long been known that during the thermal processing of amorphous ice, small molecules present at trace amounts in the ice diffuse during its restructuring, leading to segregation and desorption processes \citep{Collings04}; PAHs have been shown to aggregate during ice heating since their adsorption energy is too high for desorption to occur \citep{Lignell15,MichoulierAdsorption}.
Recent studies on reactivity in interstellar ice analogues have highlighted the important role played by the structure of ice on diffusion of reactants and production rates of interstellar complex organic molecules (iCOMs, \citet{Ghesquiere18}).
To date, PAH photochemistry has only been studied in amorphous ice.

In this paper, we present an experimental study of photoreactions of anthracene, pyrene and coronene embedded in four different water ice structures: pASW, cASW, Ic and Ih. The results are discussed in the light of our recent theoretical studies \citep{MichoulierAdsorption,MichoulierIP,Michoulier20}.

\section{Experimental methods}

Experiments were performed in a high vacuum experimental setup consisting of a stainless steel chamber (base pressure $\sim$~10$^{-7}$~mbar) containing a CsBr substrate cooled to 10~K by a closed-cycle He cryostat.  Deionised water was subjected to multiple freeze-pump-thaw cycles under vacuum to remove dissolved gases. Rare gas (He, Ar) and water were mixed in a dosing line and injected into the chamber at a rate of 1 ml\,min$^{-1}$. PAHs (anthracene, 99~\%; pyrene, 99~\%; and coronene, 99~\%, Sigma-Aldrich) were used without further purification and sublimated by heating powder in the oven. Prior to deposition, the coronene samples were preheated to 120~$^\circ$C for several hours to remove excess water. The deposition rate of PAHs, determined by the temperature of the sublimation oven (anthracene: 75~$^\circ$C, pyrene: 75~$^\circ$C, and coronene: 180~$^\circ$C), was constant for all experiments. PAHs and water:rare gas mixtures were co-deposited onto a CsBr substrate held at low temperature. 
Ices of different structures were obtained by varying the temperature of the CsBr substrate during deposition and the nature of the rare gas (to avoid it sticking): 15~K, He (pASW); 120~K, Ar (cASW); 150~K, Ar (Ic); and 175~K, Ar (Ih). Depositions lasted for between 75 and 90 minutes, depending on the PAH and on the ice structure, and deposited ices were subsequently cooled to 10~K. Infrared spectra were measured in transmission mode using a Bruker 70~V Fourier transform infrared (FTIR) spectrometer with a DTGS detector (4000--500 cm$^{-1}$, 0.5 cm$^{-1}$ resolution, each spectrum averaged over approximately 100 scans). All spectra were measured at 10~K. Ices were irradiated at 10~K with a mercury lamp for a total of 150~minutes ($>$~235~nm, average power: 150~mW, total fluence: 0.72~J\,m$^{-2}$), with IR spectra measured after various lengths of irradiation time.

\section{Results and discussion}

\begin{figure}[!b]
	\includegraphics[width=\columnwidth]{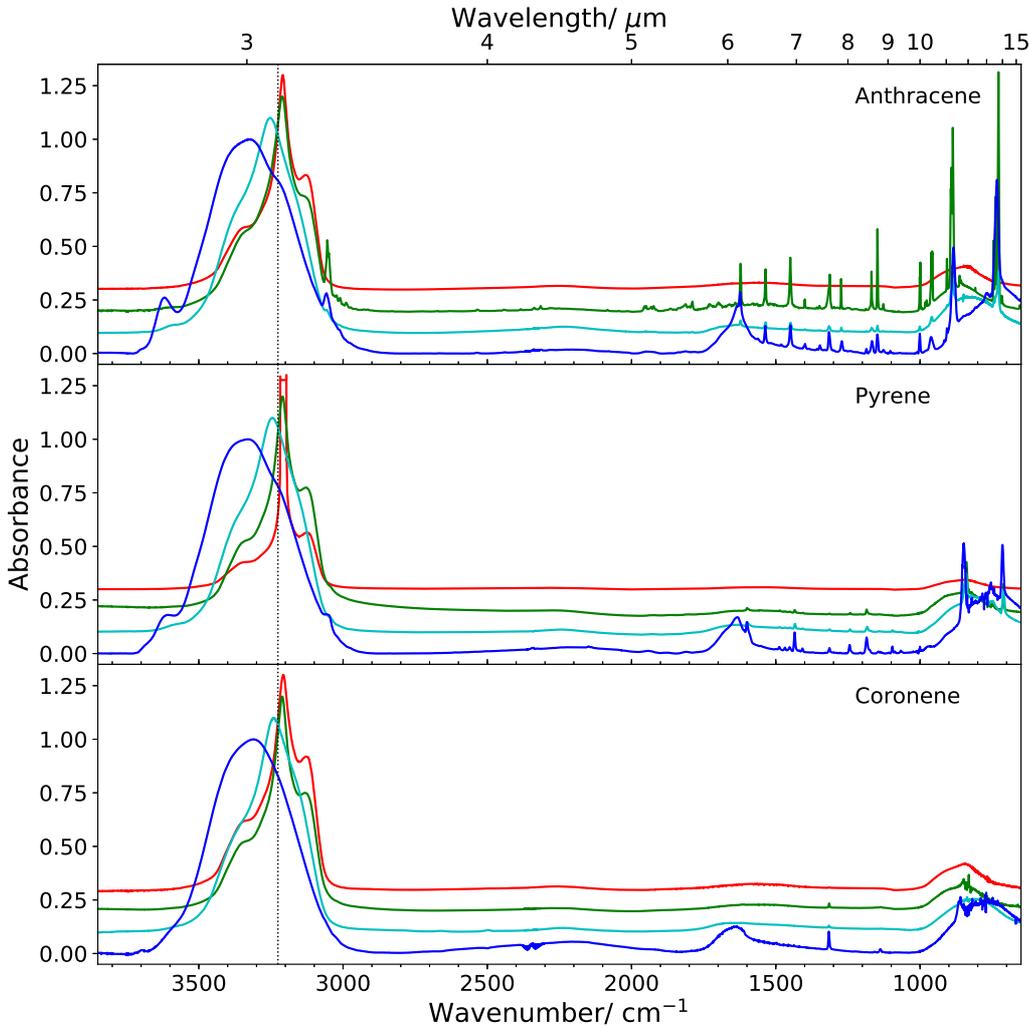}
    \caption{Spectra of all PAH:ice mixtures upon deposition. Lower to upper panels: Coronene, pyrene, anthracene; in ice structures: pASW, blue; cASW, cyan; Ic, green; Ih, red. Some data are baselined for ease of comparison, and all data are normalised to the peak of the 3.1~$\mu$m OH stretching band (marked with a dotted line) then offset by multiples of 0.1 on the absorbance scale.}
    \label{fig:spectra}
\end{figure}

Figure~\ref{fig:spectra} presents the deposition spectra of all  PAH:water ice mixtures. Data for coronene, pyrene, and anthracene are plotted in the lower, middle, and upper panels, respectively. For each PAH, depositions in pASW (blue), cASW (cyan), Ic (green), and Ih (red) ices are shown. The structure of the deposited ice can be discerned in the profile of the 3.1~$\mu$m feature, which is wide and featureless for ASW, but becomes narrower and more peaked for crystalline ices. For all PAH:water ice depositions, characteristic absorption bands of the PAHs are visible, most notably in the 1600 -- 600~cm$^{-1}$ region. This is also the spectral region where OH-bearing photoproducts of the irradiated PAH:water ice mixtures are identified.

It should be noted that, due to the nature of the deposition method, PAH:water concentrations were constrained by deposition conditions for both water and the PAH species, could not be determined in the gas phase, and had to be estimated based upon analysis of the absorption band strengths in the deposited ice spectra \citep{Gerakines95,Bouwman11,DeBarros17}.
This is an inherently uncertain method due to the influence of structure (and especially mixture) upon the relative band strengths of the different components in the solid phase. Based on this approach, anthracene and pyrene were determined to be present at approximately 1-2~\% (apart from anthracene in Ic, $\sim$~9~\%) while coronene was an order of magnitude less concentrated ($\sim$~0.1~\%) In all cases, in Ih, the concentration is likely higher as, due to the high deposition temperature of Ih (175~K), the sticking efficiency of H$_2$O is lowered and the initial PAH:H$_2$O ratio is altered. These values are summarised in Table~\ref{tab:conc}.
These differences in concentration can be observed, for example, in the pASW mixture spectra in Figure~\ref{fig:spectra}. In the case of anthracene and pyrene ($\sim$~1~\%), a clear absorption feature can be observed in the blue wing of the 3.1~$\mu$m stretching band of the water ice ($\sim$~3620~cm$^{-1}$), while it is not immediately visible in the case of coronene ($\sim$~0.1~\%). We will return later to a discussion of the 3.1~$\mu$m feature and what we can discern about ice mixture structure from this band.
It is well-established that PAH reactivity can be influenced by its relative concentration in pASW \citep{Cuyulle14,Cook15,DeBarros17}.
All our depositions, having a PAH:water concentration greater than 1:1000 (varying from 1:1000 to 3:100), fall in the region where ionisation efficiency is limited to 15~\%. Thus, this factor has not been taken into account during any comparative discussion between PAHs.
 
 \begin{table}[]
     \centering
     \begin{tabular}{c|cccc}
          PAH & pASW & cASW & Ic & Ih \\
          \hline
          Anthracene & 3 & 1 & 9 & >3 \\
          Pyrene & 1 & 1 & 1 & >1 \\
          Coronene & 0.3 & 0.1 & 0.1 & >0.3
     \end{tabular}
     \caption{Concentrations of PAH in water ices of different structures (in percent), as derived from the band areas of PAH and water in each deposition spectrum in Figure~\ref{fig:spectra}.}
     \label{tab:conc}
 \end{table}

\begin{figure}
	\includegraphics[width=\columnwidth]{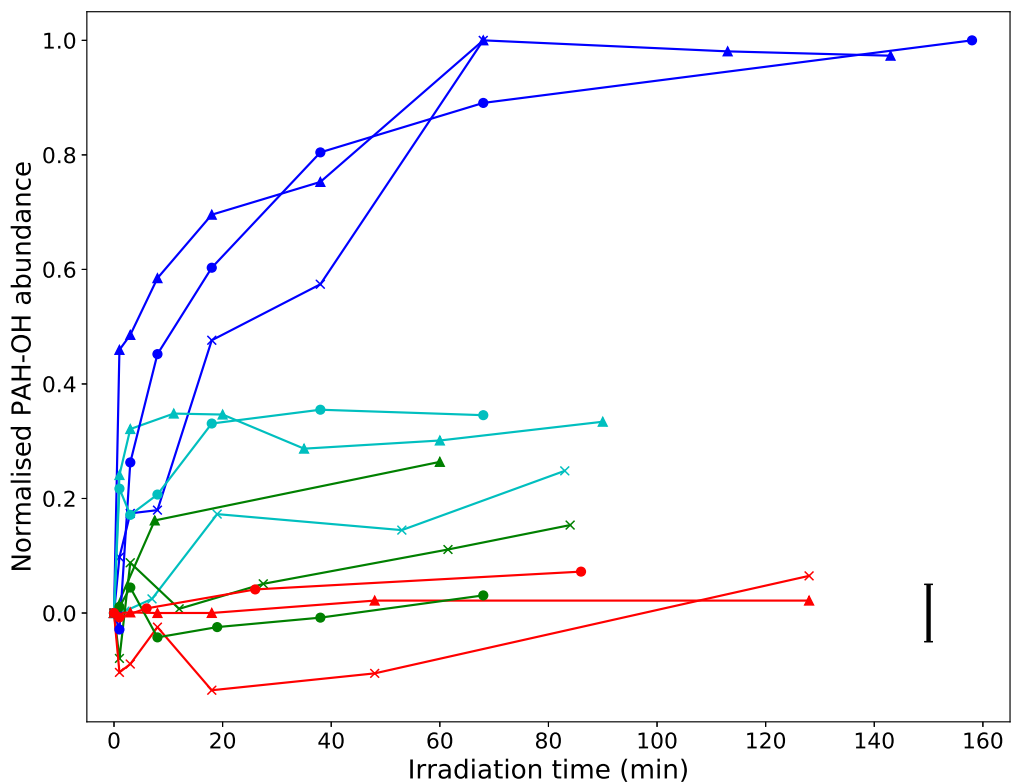}
    \caption{Normalised PAH-OH photoproduct of all PAH:water ice mixtures. PAHs: coronene, circles; pyrene, triangles; anthracene, crosses. Ices: pASW, blue; cASW, cyan; Ic, green; Ih, red. All values normalised to maximum production of PAH-OH photoproduct for each PAH. Estimated uncertainty on each value is $<$~10~\% as indicated by the error bar in the figure.}
    \label{fig:norm_photoprod}
\end{figure}

After UV irradiation of each ice ($\lambda$ > 235~nm), we searched for the appearance of absorption features characteristic of known photoproducts: hydroxy-PAHs (PAH-OH), with an -OH group replacing one of the hydrogen atoms, as well as PAH fragmentation products CO and CO$_2$.
Figure~\ref{fig:norm_photoprod} illustrates the kinetics of formation of OH-bearing photoproducts versus irradiation time for the three PAHs studied here (coronene - circles, pyrene - triangles, anthracene - crosses) in all four ice structures (pASW - blue, cASW - cyan, Ic - green, and Ih - red). The normalised PAH-OH abundance at each datapoint was calculated by taking the area under the band for PAH-OH IR spectral features located at 1208, 1145, and 1292~cm$^{-1}$ for coronene-OH \citep{DeBarros17}, pyrene-OH \citep{Cook15}, and anthracene-OH \citep{Cook15}, respectively, normalised to the maximum photo-yield of each (i.e. in pASW)). 
If we first consider pASW (blue traces), where these OH-bearing photoproducts have previously been identified upon both ``soft'' \citep{Guennoun11a,Guennoun11b} and more energetic \citep{Bouwman11,Cook15,DeBarros17} UV irradiation, we observe that, for all three PAHs, the production of hydroxy-PAH increases with irradiation time. The abundance of hydroxy-PAHs is highest for all species in the pASW ice mixtures, where they begin to form immediately (\emph{i.e.} within the first minute of irradiation) and increase rapidly, reaching a plateau after approximately one hour of irradiation.

It is also clear that this pattern is not repeated for irradiation of the other PAH:water ice structures. In cASW (cyan traces), the yield of hydroxy-PAH is reduced to around one third of that in pASW, although the production also both begins and rises rapidly, reaching a plateau earlier than in pASW. For crystalline ices, there is approximately no production of hydroxy-PAHs, with some of the data even appearing to be negative. This is both an experimental and a data analysis issue, due to fits being performed on very low intensity peaks with minor variation from one measurement to the next due to the repositioning of the experimental chamber in the spectrometer beam after each UV irradiation slightly modifying the beam path through the sample. We estimate the uncertainty from these factors to be up to 10~\% for trace species, as indicated by the error bar in Figure~\ref{fig:norm_photoprod}. We conclude that the production of hydroxy-PAHs in Ih samples is essentially zero, while there appears to be production of hydroxy-pyrene and probably hydroxy-anthracene ($\sim$~25~\% and <~15~\% that in pASW, respectively) but not hydroxy-coronene in Ic samples.

\begin{figure}
    \includegraphics[width=\columnwidth]{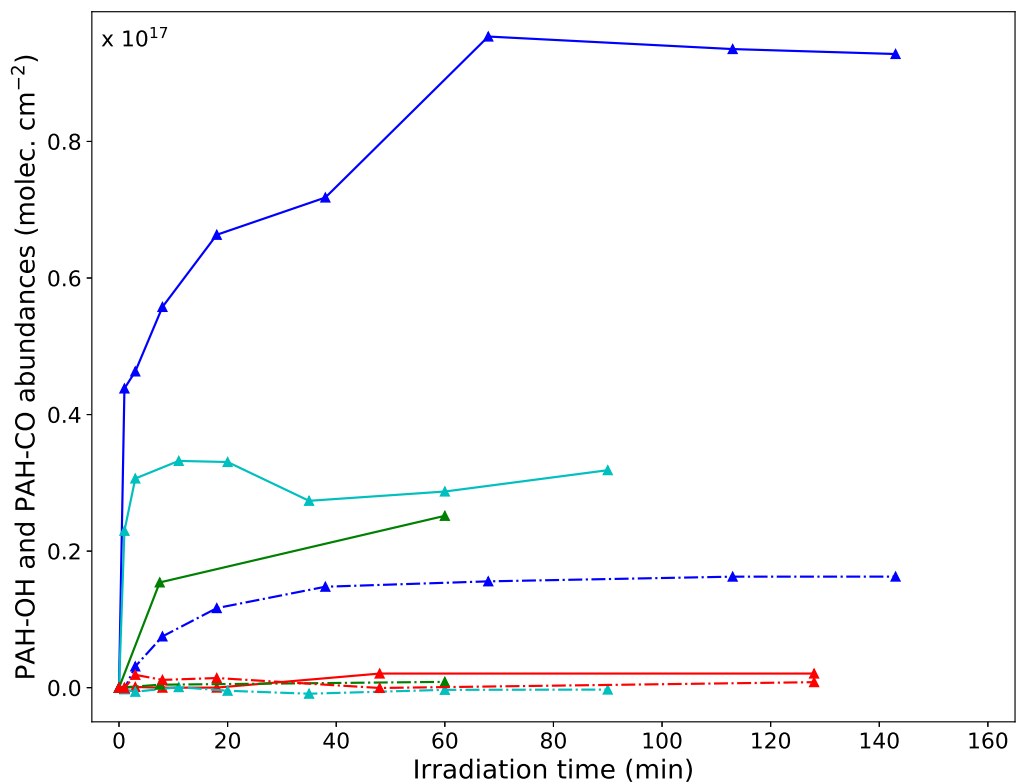}
    \caption{Oxygenated photoproducts produced upon irradiation of pyrene:water ices (pASW - blue, cASW - cyan, Ic - green, Ih - red). Pyrene-OH production is traced in filled lines, while pyrene-CO production is traced in dot-dashed lines.}
    \label{fig:pyrene_photoprod}
\end{figure}

Figure~\ref{fig:pyrene_photoprod} presents the absolute abundances of oxygen-bearing photoproducts for pyrene, pyrene-OH and the ketonic pyrene-CO. In pASW, photoproducts bearing an OH group appear almost immediately, while the formation of pyrene bearing one C=O bond (\emph{i.e.} corresponding to subsequent H-loss) begins later, in agreement with previous studies \citep[e.g.][]{Cook15}. Both species reach plateaus, but the maximum production of pyrene-CO is $\sim$~17~\% the maximum of pyrene-OH. It should be noted that a combination of low band strengths and low intensities meant it was not possible to perform the same analysis for anthracene-CO and coronene-CO in these experiments. In previous studies, CO-bearing photoproducts have been identified for coronene \citep{DeBarros17}, anthracene \citep{Ashbourn07,Cook15}, and pyrene \citep{Cook15}.
Low abundance could also account to some extent for the non-observation of pyrene-CO in cASW, Ic, and Ih (although it is not expected to form in Ih since no pyrene-OH was produced).

The two main results from our data on oxygen-bearing photoproducts of PAHs -- that photo-oxidation has a plateau and that it is favoured in amorphous ices -- both suggest that this reactivity is driven by the presence of PAH molecules in certain, favourable sites in the ice mixtures. These sites appear to be more prevalent in amorphous ices, especially porous amorphous ice, and once reactivity occurs at these sites, no more hydroxy-PAHs are formed. We have already investigated the orientation-dependence of PAH:water interactions in both water clusters \citep{Noble17} and in the adsorption characteristics of PAHs on ice surfaces \citep{MichoulierAdsorption}. In \citet{Noble17}, we showed experimentally that the presence of an ice was not a prerequisite for the formation of hydroxy-PAHs, but that we could form coronene-OH upon soft UV ($\lambda$ > 235~nm) irradiation of clusters of a PAH molecule with up to four water molecules isolated in a cryogenic matrix. Ab initio calculations suggested that favourable geometries for small clusters were for the water molecules to form an aggregate bound to the edge of the PAH. Upon UV-initiated electronic excitation, this cluster geometry gave rise to a charge transfer state, via which we postulated H transfer from a C-H of coronene to the water could be the first step towards the formation of coronene-OH.
In a later theoretical study, we also showed that the ionisation potential of a PAH was sensitive to its interaction geometry with an ice surface, with PAHs preferentially bonding to dangling OH bonds at the ice surface \citep{MichoulierAdsorption} and the interaction of the C-H bonds at a PAH edge with O in the water ice surface lowering the ionisation potential of the PAH molecule by up to 0.8~eV \citep{MichoulierIP}. All of these results highlight the key role of ice structure in its reactivity. 

\begin{figure}
    \includegraphics[width=\columnwidth]{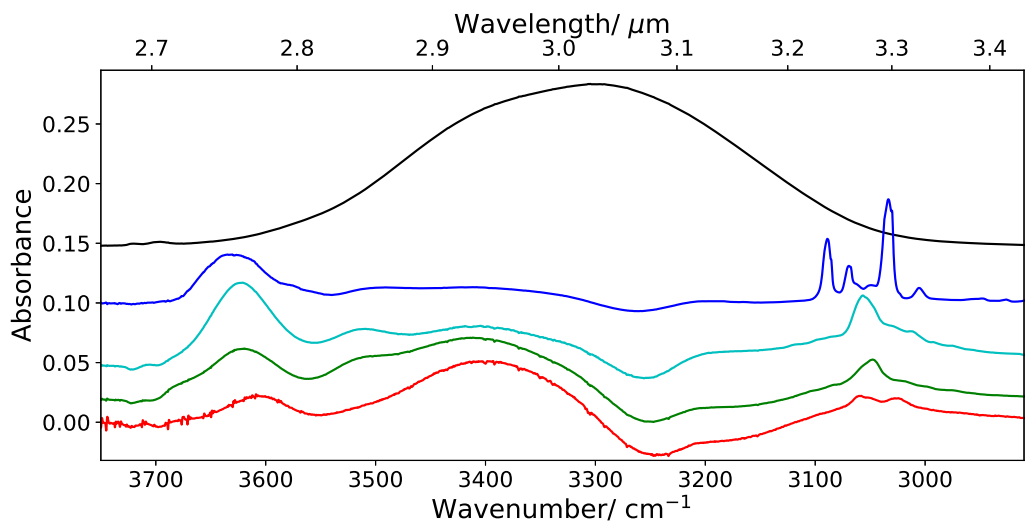}
    \caption{Restructuring of pASW by the presence of aromatic molecules. The spectrum of pure pASW (black trace) is subtracted from that of aromatic:water ice mixtures to derive difference spectra, presented in coloured traces, from lower to upper: coronene, red; pyrene, green; anthracene, cyan; and the single aromatic ring benzene in blue, for completeness. Spectra correspond to pASW depositions presented in Figure~\ref{fig:spectra} and are extended versions of those published in \citet{Michoulier20}. The spectra of pASW, coronene:pASW and anthracene:pASW are multiplied by factors 0.2, 2, and 0.5, respectively, to ease visual comparison.}
    \label{fig:restructuring}
\end{figure}

When we discuss reactivity in ices, we typically concentrate on the influence of the ice upon the molecule dispersed within the ice, but in this case it is also worth considering the influence of the relatively large PAH molecules upon the ice structure. To do this we can compare the structure of the PAH:pASW ice mixtures to a pure pASW deposited under equivalent experimental conditions. In Figure~\ref{fig:restructuring} we present the 3.1~$\mu$m stretching absorption band of a pure pASW sample deposited in our experimental setup (black trace). The other traces correspond to difference spectra, \emph{i.e.} the spectrum of a PAH:water ice mixture minus this (scaled) pure pASW spectrum (coronene: red, pyrene: green, anthracene: cyan, and, for completeness, the single aromatic ring, benzene: blue).  The presence of aromatic molecules can be discerned by the presence of bands in the $\sim$~3100--3000~cm$^{-1}$ region, corresponding to C-H stretches of the aromatic molecules, as well as a wide feature at $\sim$~3650~cm$^{-1}$ attributed to dangling OH bonds at the surface of the ice in interaction with the aromatic molecule. We recently published a dedicated combined experimental-theoretical study on the influence of aromatics on the position and width of this feature \citep{Michoulier20}. However, in Figure~\ref{fig:restructuring} we can observe that the bulk ice is also impacted by the presence of PAH molecules.  We identify the restructuring of the bulk ice compared to pure pASW by the presence of an increased absorbance at 3400~cm$^{-1}$ and a corresponding decreased absorbance at 3250~cm$^{-1}$ upon addition of PAH. In pASW, oscillators in the red wing ($\sim$~3250~cm$^{-1}$) are attributed to water molecules which are fully tetrahedrally bound in the water matrix, while those in the blue wing ($\sim$~3400~cm$^{-1}$) are identified as triply-bound species at the surface \citep{Devlin95,Rowland95,Buch97}. Thus, the presence of aromatic molecules in pASW seems to disrupt the ice structure, forcing it into a less ordered ice than in pure pASW. The very presence of PAH molecules at fractions of a few percent within the ice sample tends to lead to a less structured, more amorphous solid. This would seem to suggest that the presence of the PAH as a trace impurity in the ice stabilises the metastable phase, in line with the results of \citet{Lignell15} who observed that crystallisation of a mixed pyrene:ASW ice at 140~K occurs significantly more slowly than for a pure water ice.
\\

\begin{figure*}
    \includegraphics[width=\textwidth]{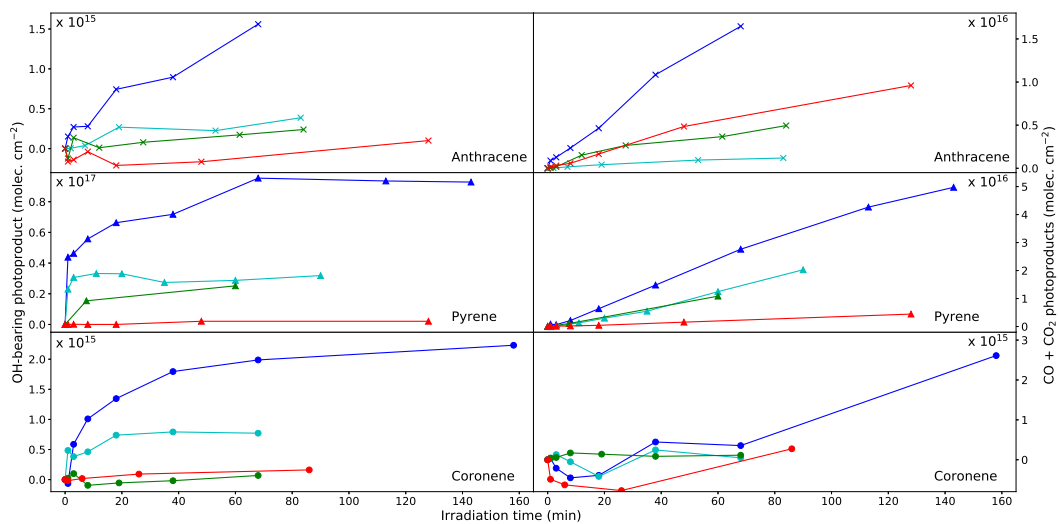}
    \caption{Hydroxy-PAH photoproducts (left) and photofragments (CO and CO$_2$, right) produced upon UV irradiation of PAH:water ice mixtures consisting of coronene (lower panels), pyrene (middle panels), and anthracene (upper panels) in pASW (blue), cASW (cyan), Ic (green) and Ih (red) ices.}
    \label{fig:all_photoprod}
\end{figure*}

As well as hydroxy-PAH reaction products (and their subsequent dehydrogenation products), UV irradiation of PAHs and PAH:water ice mixtures gives rise to photofragmentation of the PAHs \citep{Cook15,DeBarros17}. Figure~\ref{fig:all_photoprod} presents the photoproduct formation versus irradiation time for all PAH:water ice mixtures, with the hydroxy-PAHs presented in the left-hand panels  and the sum of the fragments CO and CO$_2$ in the right-hand panels (lower to upper panels: coronene, pyrene, anthracene). 
The total quantity of fragments was calculated from the absorbance of CO and CO$_2$ measured at 2149 and 2343~cm$^{-1}$ in the FTIR spectra \citep{Gerakines95}.  
First, we can observe that erosion of PAHs occurs on a relatively similar scale to their photoreactivity in pASW (blue traces), with a maximum difference of only an order of magnitude observed between the total number of OH-bearing photoproducts and total number of fragments generated by the end of each experiment.
It is interesting to note that the yield of photofragments is highest in pASW mixtures for all PAHs, in line with the yield of hydroxy-PAH photoproducts.

For none of the PAH:water ice mixtures does the kinetics of fragmentation follow that of hydroxy-PAH formation. The production of CO and CO$_2$ is approximately linear with time in all mixtures of anthracene and pyrene (upper and middle right-hand panels), with fragmentation seemingly directly linked to irradiation time. In coronene (lower right-hand panel), there is approximately no fragmentation observed in any mixture up to $\sim$~70 minutes of irradiation, with fragmentation beginning after this in pASW only. Again, a combination of experimental and data analysis difficulties can sometimes lead to a derived value of below zero in the production of photofragments for coronene. In these cases we assume no photofragmentation has occurred.
It would appear that the smaller the PAH (in this limited study of only three species), the higher the proportion of fragmentation in pASW. For anthracene, the rate of photofragmentation in pASW is approximately an order of magnitude higher than its photoreactivity, even at early times, while in the case of coronene, fragmentation appears to begin concurrently with the plateau of hydroxy-PAH production, and only in pASW.
It has long been known that the larger a PAH, the more resistant it is to photofragmentation \citep{Zhen15,Zhen16}. A recent experimental study has confirmed that, for isolated PAH cations subject  to  higher  energy  (VUV)  photons,  photoionisation dominates over photofragmentation for larger PAHs ($\sim$30--48 C), confirming that the size effect can even compensate for a double charge \citep{Joblin20}.
Is the possible size-dependence in our PAH:ice mixtures indicative of more efficient energy dissipation in the case of coronene, or does the coronene:water ice interaction also play a role? It is not possible to tease out an answer to such a complex question in a study with this limited size range of PAHs.

Considering the impact of ice structure, it is difficult to discern any pattern for coronene, as there is almost no fragmentation in any of the ices. In the case of pyrene, the relative production of fragments follows the order observed for the production of hydroxy-pyrene, i.e. maximum fragmentation in pASW, approximately no fragmentation in Ih, and an intermediate fragmentation observed for cASW and Ic. For anthracene, the pattern is very different, with a maximum fragmentation in pASW, but an intermediate fragmentation observed for Ih, with lower fragmentation in cASW and Ic. It would therefore appear that, while PAHs seem to be less sheltered from fragmentation in pASW, the influence of ice structure upon PAH fragmentation dynamics is a much more subtle effect than its influence upon hydroxy-PAH formation. It could be that PAH size and ice structure are competing factors in the dissipation of energy within these systems, giving rise to unpredictable behaviour. It could also be that PAH concentration is much more critical to fragmentation than to hydroxy-PAH formation. This will be more fully tested in a future set of dedicated experiments.

\section{Astrophysical implications}

This experimental study has implications for both the astrophysical formation and observation of hydroxy-PAHs. The first key conclusion is that the presence of larger molecules such as PAHs within an amorphous ice can modify the profile of the 3.1~$\mu$m OH stretching band, and thus this could potentially be used to search for the presence of large molecules (PAHs) in the solid phase. However, more experimental work would first be required to test the impact of larger PAHs on ice structure, and there are many other factors influencing the 3.1~$\mu$m band profile (such as grain shape and size effects in the red wing, and the presence of absorption features of species such as NH$_3$ and CH$_3$OH) that would need to be taken into account.

It would appear from our results that PAHs react rapidly in pASW ice to form oxygenated photoproducts but are less likely to photooxidise or to photofragment in more ordered ice structures.
This could be very useful in constraining the formation pathways and timescales of oxygen-bearing aromatic molecules, such as those detected in meteorites. Oxygen-bearing aromatic molecules have been identified in carbonaceous chondrite meteorites, such as Murchison  \citep[e.g.][]{Krishnamurthy92}, although there is some debate as to their formation process. As proposed by \citet{Bernstein99}, the oxygen-bearing PAH species observed in these meteorites could have been formed by UV irradiation of PAHs in water ice, which also has the merit of explaining isotopic ratios in these organic molecules. Our results indicate that, if this reactivity occurred in the ice and not during later pyrolytic conversion, it is most likely to have occurred in amorphous rather than crystalline ice, i.e. it would most likely be in cold, dense regions where the ice is in an amorphous structure, although the UV flux is relatively low in such environments. Once the ice becomes more ordered, for example in warmer regions, any PAHs trapped in the ice would no longer react upon irradiation, but rather fragment or dissipate the energy into the icy mantle.
The heating of ices, as well as shocks or photodesorption, could release oxygenated PAHs into the gas phase following their formation in icy mantles, rendering their gas phase detection possible. Oxygenated PAHs that are not released from the grain could be integrated into small body and planet formation, giving rise to targets for studies of cometary bodies to constrain earlier ice conditions.
%


In summary, our results show that the production of oxygenated PAHs is most efficient in porous amorphous water ice, where dangling H bonds are available in pores, confirming that the reaction occurs with water in favourable geometries, similar to small water clusters, rather than crystalline ice structures \citep{Noble17}. This is confirmed by the small IP variation induced by ice environment found in our recent calculations \citep{MichoulierIP}. This suggests that excited states rather than ionised species could play a role in the photo-reactivity of PAHs with water ice. 
This study demonstrates the importance of laboratory work to elucidate the effect of intermolecular interactions in order to better understand photochemical processes in realistic astrochemical ice environments and highlights the critical role played by molecular orientation, which itself is directed by ice structure and ice-molecule interactions. These results also give rise to the possibility of the formation of oxygenated PAH molecules in interstellar environments with low water abundance or higher temperatures. Such processes could considerably influence the inventory of aromatics in meteorites. 

\begin{acknowledgements}
We thank C.~Toubin, A.~Simon, and N.~Ben~Amor for useful discussions. 
This work has been funded by the Agence Nationale de la Recherche (ANR) project ``PARCS'' ANR-13-BS08-0005, with support from the French research network EMIE (Edifices Mol\'{e}culaires Isol\'{e}s et Environn\'{e}s, GDR 3533 of CNRS), and the French National Programme "Physique et Chimie du Milieu Interstellaire" (PCMI) of the CNRS/INSU with INC/INP, co-funded by CEA and CNES.
\end{acknowledgements}


\end{document}